\documentclass[prl, twocolumn, showkeys, amsmath, amssymb]{revtex4}

\usepackage{graphicx}

\begin{document}

\title{Selective advantage for multicellular replicative strategies:  A two-cell example}

\author{Emmanuel Tannenbaum}
\email{etannenb@gmail.google.com}
\affiliation{Ben-Gurion University of the Negev,
Be'er-Sheva, Israel 84105}

\begin{abstract}

This paper develops a quasispecies model where cells can adopt a two-cell
survival strategy.  Within this strategy, pairs of cells join together, at
which point one of the cells sacrifices its own replicative ability for
the sake of the other cell.  We develop a simplified model for the evolutionary 
dynamics of this process, allowing us to solve for 
the steady-state using standard approaches from quasispecies theory.  We find that 
our model exhibits two distinct regimes of behavior:  At low concentrations of 
limiting resource, the two-cell strategy outcompetes the single-cell survival strategy, 
while at high concentrations of limiting resource, the single-cell survival strategy 
dominates.  The single-cell survival strategy becomes disadvantageous at low concentrations 
of limiting resource because the energetic costs of maintaining reproductive and metabolic 
pathways approach, and may even exceed, the rate of energy production, leaving little excess 
energy for the purposes of replicating a new cell.  However, if the rate of energy production 
exceeds the energetic costs of maintaining metabolic pathways, then the excess energy, if 
shared among several cells, can pay for the reproductive costs of a single cell, leaving 
energy to replicate a new cell.  Associated with the two solution regimes of our model 
is a localization to delocalization transition over the portion of the genome coding for 
the multicell strategy, analogous to the error catastrophe in standard quasispecies models.  
The existence of such a transition indicates that multicellularity can emerge because natural 
selection does not act on specific cells, but rather on replicative strategies.  Within 
this framework, individual cells become the means by which replicative strategies are propagated.  
Such a framework is therefore consistent with the concept that natural selection does not 
act on individuals, but rather on populations.  

\end{abstract}


\maketitle

One of the most interesting questions under investigation in evolutionary biology
is the emergence of cooperation and multicellularity in biological systems \cite{MULTICELL, JMSMITH}
(and references therein).  While the emergence of certain types of cooperative behavior, 
such as division of labor, are reasonably well understood phenomena, the evolution of 
multicellular organisms is a more difficult question.  

With division of labor, a group of cells can more efficiently metabolize environmental
resources than if they worked alone, and so it is in each cell's replicative
interest to cooperate with other cells.  In the case of multicellular organisms, 
however, certain cells forgo their ability to replicate, so that other cells in the 
organism can survive and reproduce.  This is clearly against the replicative 
interests of the non-replicating cells, a situation that makes the strategy prone to
defections.  Indeed, defection from a multicellular survival strategy,
otherwise known as cancer, is a common phenomenon in multicellular
organisms.

Nevertheless, in certain environments, there must exist selective
pressures driving the emergence of multicellular organisms.  Perhaps one of the 
clearest demonstrations of such selective pressures is the existence of
the organism {\it Dictyostelium discoideum}, commonly known as a cellular {\it slime mold}.  
The slime mold has been the focus of considerable research (it is an NIH model organism), 
because it lives at the border between unicellular and multicellular life:  When conditions
are favorable, the slime mold exists as a collection of free-living, single-celled
organisms.  However, when the slime mold cells are stressed, say by depletion 
of some necessary resource, they respond by coalescing into a differentiated,
multicellular organism.  When conditions improve, the slime mold reproduces by
sporulation \cite{BRITTON}.

One of the interesting features of the slime mold is that, during the differentiation
process, some cells inevitably forgo replication for the sake of the multicellular
structure \cite{BRITTON}.  In this Letter, we attempt to elucidate the selective pressures driving
this behavior by considering a highly simplified model motivated by the slime mold life cycle, 
one which we believe illustrates the underlying principles involved in the emergence of multicellularity.
We emphasize, however, that this Letter does not consider the evolutionary dynamics modeling
how such behavior could have emerged in the first place.  We should also emphasize that, although
our model is motivated by the slime mold life cycle, it is believed that the ability to engage
in multicellular behavior is ubiquitous amongst single-celled organisms, and may even characterize
the organization of bacterial biofilms \cite{MULTICELL}.

For our model, we consider a population of organisms whose genomes consist of three distinct
genes (or more appropriately, genome regions):  (1)  A reproduction region, denoted $ \sigma_R $, 
coding for all the various cellular machinery involved in the growth and reproduction of the organism.  
(2)  A metabolism region, denoting $ \sigma_M $, coding for all the various cellular machinery involved in 
procuring food from the environment, and metabolizing it to release the energy required for various cellular 
processes (as in the metabolism of glucose and the storage of the energy into ATP).  (3)  A multicellular 
region, coding for the machinery necessary to implement the two-cell survival strategy.  Among the various
machinery required to implement the two-cell survival strategy is a switch that causes one of the cells
to shut off its reproductive pathways, and to devote itself to metabolizing food from the environment
for the sake of the other cell.  This part of the genome is denoted by $ \sigma_S $.

The full cellular genome is denoted by $ \sigma = \sigma_R \sigma_M \sigma_S $.
We assume that there exist master sequences, $ \sigma_{R, 0} $, $ \sigma_{M, 0} $, and
$ \sigma_{S, 0} $, corresponding to gene sequences coding for the appropriate enzymes necessary
for the proper functioning of the various systems.  In this single-fitness-peak approximation,
any mutation to these master sequences leads to the loss in function of the corresponding system.
A cell for which $ \sigma = \sigma_{R, 0} \sigma_{M, 0} \sigma_{S, 0} $ replicates via a two-cell
strategy, whereby it seeks out and joins with another cell with an identical genome.  The pathways 
encoded within $ \sigma_{S, 0} $ cause one of the cells to shut off its reproductive
pathways, and to devote its metabolic efforts to sustaining the other cell (a possible algorithm
that the switch could implement is to instruct a cell to shut off its reproductive pathways if the
reproductive pathways of the other cell is on, and to turn on its reproductive pathways if the
reproductive pathways of the other cell is off.  The only two stable solutions to this algorithm
are where one of the cells has its reproductive pathways on, while the other cell has its
reproductive pathways off.  Presumably, although the two cells join with both of their reproductive
pathways on, random fluctuations will break the symmetry and lead to collapse into an equilibrium
state).

A cell for which $ \sigma = \sigma_{R, 0} \sigma_{M, 0} \sigma_S $, $ \sigma_S \neq \sigma_{S, 0} $,
replicates independently of the other cells.  It is assumed that all other genotypes, with faulty
copies of either reproductive and metabolic pathways, do not replicate at all.

The cells metabolize a single external resource, which provides both the energy and the raw 
materials for all the cells' needs.  If we let the basic unit of energy be the amount
of energy released by metabolism of a set quantity of resource, then up to a conversion factor it is
possible to measure all energy and accumulation changes in terms of the resource itself.  Of course,
because only that quantity of resource that has been metabolized has provided the cell with energy and
raw materials, our basic measurement unit becomes the quantity of metabolized resource.

It is assumed that resource is metabolized by each cell via a two-step process:  (1)  A binding step, whereby the 
resource binds to certain receptors, which then pass on the resource for metabolism.  (2)  A metabolism step, whereby
the resource bound the receptors is then metabolized.  Assuming each of the steps is an elementary
reaction, we obtain a metabolism rate $ r(c) $ of the Michaelis-Menten form $ \alpha c/(1 + \beta c) $,
where $ c $ denotes the concentration of resource in the environment.  Note that this form of the 
metabolism rate has the property that it reaches a maximal value as the concentration of external resource 
becomes infinite.  This makes sense, since a cell cannot metabolize an external resource at arbitrarily
high rates.  It should be noted, however, that our expression for $ r(c) $ is not the only one that exhibits 
this saturation property, but it is one of the simplest expressions possible.

In order to replicate a cell, the various cellular systems must be replicated.  Each system 
has an associated {\it build cost} (measured in units of metabolized resource).  Thus, if $ \rho_R $, $ \rho_M $,
and $ \rho_S $ denote the build costs of the reproductive, metabolic, and two-cell pathways, respectively, 
then the total cost required to build a new cell replicating via the single-cell strategy is given by 
$ \rho_R + \rho_M $, while the total cost required to build a new cell replicating via the two-cell strategy
is given by $ \rho_R + \rho_M + \rho_S $.  

In addition to the build costs for the various systems, each system has an associated {\it fixed cost},
corresponding to the energy and resources required to maintain system function.  These fixed costs
arise because the various components of the cellular systems have intrinsic decay rates (protein 
degredation, auto-hydrolysis of mRNAs, etc.), and in the case of switches that have to respond
to changes in the external environment or the internal states of the cell, there is a minimal rate
of energy consumption associated with measuring ambient conditions.

There is also an {\it operating cost} associated with each subsystem, corresponding to energy and resource 
costs associated with carrying out a given system task.  For example, the replication machinery consumes  
energy in order to process a certain amount of metabolized resource toward the construction of a new cell.  
The metabolic pathways require energy to break-down the external resource (in chemistry, such costs are known as
activation barriers).  

Let $ \omega_R $ denote the cost of replication per unit of metabolized resource incorporated 
into a new cell, and let $ \omega_M $ denote the cost of metabolizing one unit of resource.  Then 
for a cell replicating via the single-cell strategy, the total amount of resource that must be 
metabolized is given by, $ \rho \equiv = (1 + \omega_M) (\rho_R + \rho_M) $.  The net rate of 
energy production is given by $ (1 - \omega_M) r(c) $.  Since replication and metabolism consume 
energy at a rate given by $ \dot{\rho} \equiv \dot{\rho}_R + \dot{\rho}_M $, the net rate of energy 
accumulation is given by, $ (1 - \omega_M) \alpha c/(1 + \beta c) - \dot{\rho}_R - \dot{\rho}_M $.
The replication time is therefore given by,
\begin{equation}
\tau_{rep} = \frac{\rho}{(1 - \omega_M) r(c) - \dot{\rho}}
\end{equation}
yielding a first-order growth-rate constant of
\begin{equation}
\kappa_1(c) = \frac{1}{\tau_{rep}} = \frac{(1 - \omega_M) r(c) - \dot{\rho}}{\rho}
\end{equation}

For the two-cell replication strategy, the cell that is replicating in the cell-pair must
accumulate a total of $ (1 + \omega_M) (\rho_R + \rho_M + \rho_S) = \rho + \Delta \rho $
of metabolized resource.  The net rate of energy production from both cells is given by
$ 2 (1 - \omega_M) r(c) $.  Since only one of the cells in the replicating cell-pair
has active reproductive pathways, the total energy consumption rate is given by
$ 2 (\dot{\rho} - \Delta \dot{\rho}) = \dot{\rho}_R + 2 \dot{\rho}_M + 2 \dot{\rho}_S $,
where $ \Delta \dot{\rho} \equiv (1/2) (\dot{\rho}_R - 2 \dot{\rho}_S) $.  Therefore,
the replication time is given by,
\begin{equation}
\tau_{rep} = \frac{1}{2} \frac{\rho + \Delta \rho}{(1 - \omega_M) r(c) - \dot{\rho} + \Delta \dot{\rho}}
\end{equation}
yielding a first-order growth-rate constant of
\begin{equation}
\kappa_2(c) = 2 \frac{(1 - \omega_M) r(c) - \dot{\rho} + \Delta \dot{\rho}}{\rho + \Delta \rho}
\end{equation}
We should note that we are implicitly assuming in this derivation that the amount of time
it takes for two cells to find each other and combine is negligible compared to the replication
time.  We are also assuming that the costs associated with transporting metabolized resource
from one cell to another is negligible.  Finally, we are also assuming that the reproductive
pathways can process the metabolized resource as fast as it is produced.

\begin{figure}
\includegraphics[width = 0.9\linewidth, angle = 0]{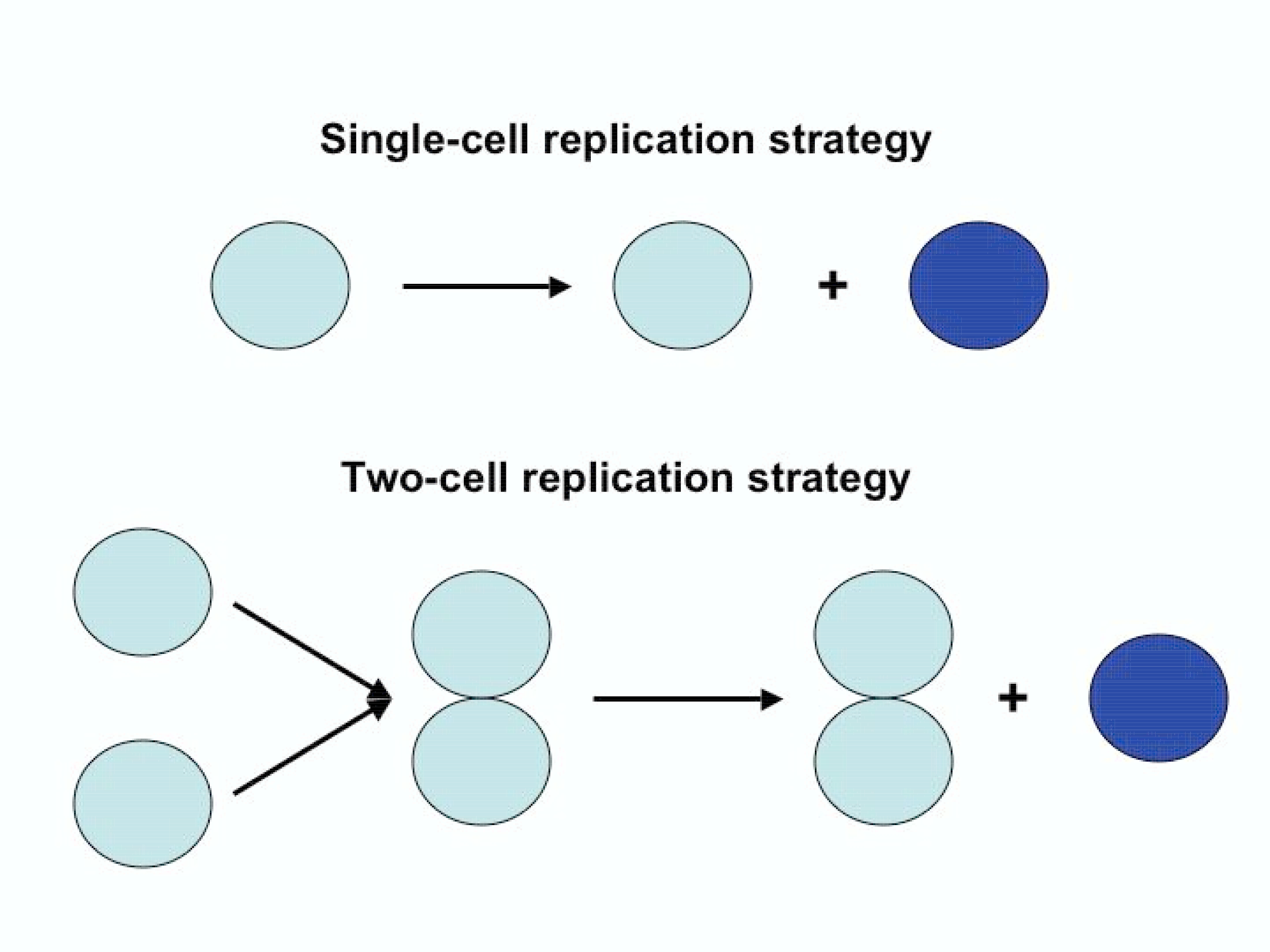}
\caption{Comparison of the single-cell and the two-cell replication strategies.}
\end{figure}

We let $ n_1 $ denote the number of organisms with the
single-cell genome.  Because we are neglecting the time it
takes for two organisms replicating via the two-cell strategy
to find each other and to combine, we may assume that all such
cells exist in the two-cell state.  We therefore define $ n_2 $
to be the number of such cell-pairs in the system.  Then
define the total population of cells $ n = n_1 + 2 n_2 $,
and population fractions $ x_1 = n_1/n $ and
$ x_2 = 1 - x_1 = 2 n_2/n $.

We also assume that cells may generate mutated daughter cells as a result of point-mutations 
during replication.  For simplicity, we assume that replication of the master sequences 
$ \sigma_{R, 0} $ and $ \sigma_{M, 0} $ is error-free, so that we do not need to consider 
cells with faulty reproduction or metabolic pathways (this situation can be created by 
assuming that the portions of the genomes coding for reproduction and metabolism are short, 
so the probability of mutations occurring in these regions is negligible).  However, 
we assume that the per-base replication error probability in $ \sigma_S $ is given by 
$ \epsilon $.  We let $ L $ denote the length of $ \sigma_S $, and define $ \mu = 
L \epsilon $.  We then consider the infinite sequence length limit, while holding 
$ \mu $ constant.  In this limit, the probability of correctly replicating $ \sigma_S $ is
given by $ p_S = e^{-\mu} $.  We then have,
\begin{eqnarray}
&   &
\frac{d x_1}{dt} = (\kappa_1(c) - \bar{\kappa}(t)) x_1 + \frac{1}{2} \kappa_2(c) (1 - p_S) x_2 \\
&   &
\frac{d x_2}{dt} = (\frac{1}{2} \kappa_2(c) p_S - \bar{\kappa}(t)) x_2 \\
&   &
\frac{d n}{dt} = \bar{\kappa}(t) n
\end{eqnarray}
where $ \bar{\kappa}(t) = \kappa_1(c) x_1 + \frac{1}{2} \kappa_2(c) x_2 $.

The above population fractions will evolve to a steady-state \cite{QUASREV}, whose properties
we can readily determine:  The condition that $ d x_2/dt = 0 $ at steady-state 
implies that either $ x_2 = 0 $ or $ \bar{\kappa}(t = \infty) = \frac{1}{2} \kappa_2(c) p_S $.  
If $ x_2 = 0 $, then $ d x_1/dt = 0 $ implies that $ \bar{\kappa}(t = \infty) = \kappa_1(c) $.  

For a steady-state to be stable to perturbations, we must have $ \bar{\kappa}(t = \infty) 
\geq \kappa_1(c), \frac{1}{2} \kappa_2(c) p_S $.  Therefore, at steady-state we have,
$ \bar{\kappa}(t = \infty) = \max\{\frac{1}{2} \kappa_2(c) p_S, \kappa_1(c)\} $.
Using the formulas for $ (1/2) \kappa_2(c) p_S $ and $ \kappa_1(c) $, and assuming
that $ \Delta \dot{\rho} > 0 $, we have that
\begin{eqnarray}
&   &
\frac{1}{2} \kappa_2(c) p_S > \kappa_1(c), \mbox{ } x_2 > 0, \mbox{ if $ 0 \leq r(c) < r(c)_{=} $} \\
&   &
\frac{1}{2} \kappa_2(c) p_S < \kappa_1(c), \mbox{ } x_2 = 0, \mbox{ if $ r(c) >  r(c)_{=} $}
\end{eqnarray}
where,
\begin{equation}
r(c)_{=} = \frac{\dot{\rho}}{1 - \omega_M} \frac{1 - p_S \frac{\dot{\rho} - \Delta{\dot{\rho}}}{\dot{\rho}}
\frac{\rho}{\rho + \Delta \rho}}{1 - p_S \frac{\rho}{\rho + \Delta \rho}}
\end{equation}

Let $ z_{1, l} $ denote the fraction of the population whose genome 
$ \sigma_{R, 0} \sigma_{M, 0} \sigma_S $ is such that $ D_H(\sigma_S, \sigma_{S, 0}) = l $, 
where $ l > 0 $.  Then, using similar techniques to those found in \cite{QUASREV}, it is possible to show that,
\begin{equation}
\frac{d z_{1, l}}{dt} =
\frac{1}{2} \kappa_2(c) x_2 \frac{\mu^{l}}{l!} e^{-\mu} +
\kappa_1(c) e^{-\mu} \sum_{l' = 0}^{l-1} \frac{\mu^{l'}}{l'!} z_{1, l-l'}
- \bar{\kappa}(t) z_{1, l}
\end{equation}
Defining the localization length $ \langle l \rangle $ via,
\begin{equation}
\langle l \rangle_S = \sum_{l = 1}^{\infty} l z_{1, l}
\end{equation}
then at steady-state,
\begin{equation}
\langle l \rangle_S = \mu \frac{\bar{\kappa}(t = \infty)}{\bar{\kappa}(t = \infty) - \kappa_1(c)}
\end{equation}
which is finite as long as $ \bar{\kappa}(t = \infty) = \frac{1}{2} \kappa_2(c) p_S > \kappa_1(c) $,
and $ \infty $ otherwise.

In other words, once the selective advantage for replicating via the two-cell survival strategy
disappears, the portion of the genome coding for this strategy undergoes a localization
to delocalization transition, analogous to the error catastrophe (it is also similar to
a phenomenon known as ``survival of the flattest'') \cite{QUASREV, WILKE}.

\begin{figure}
\includegraphics[width = 0.9\linewidth, angle = 0]{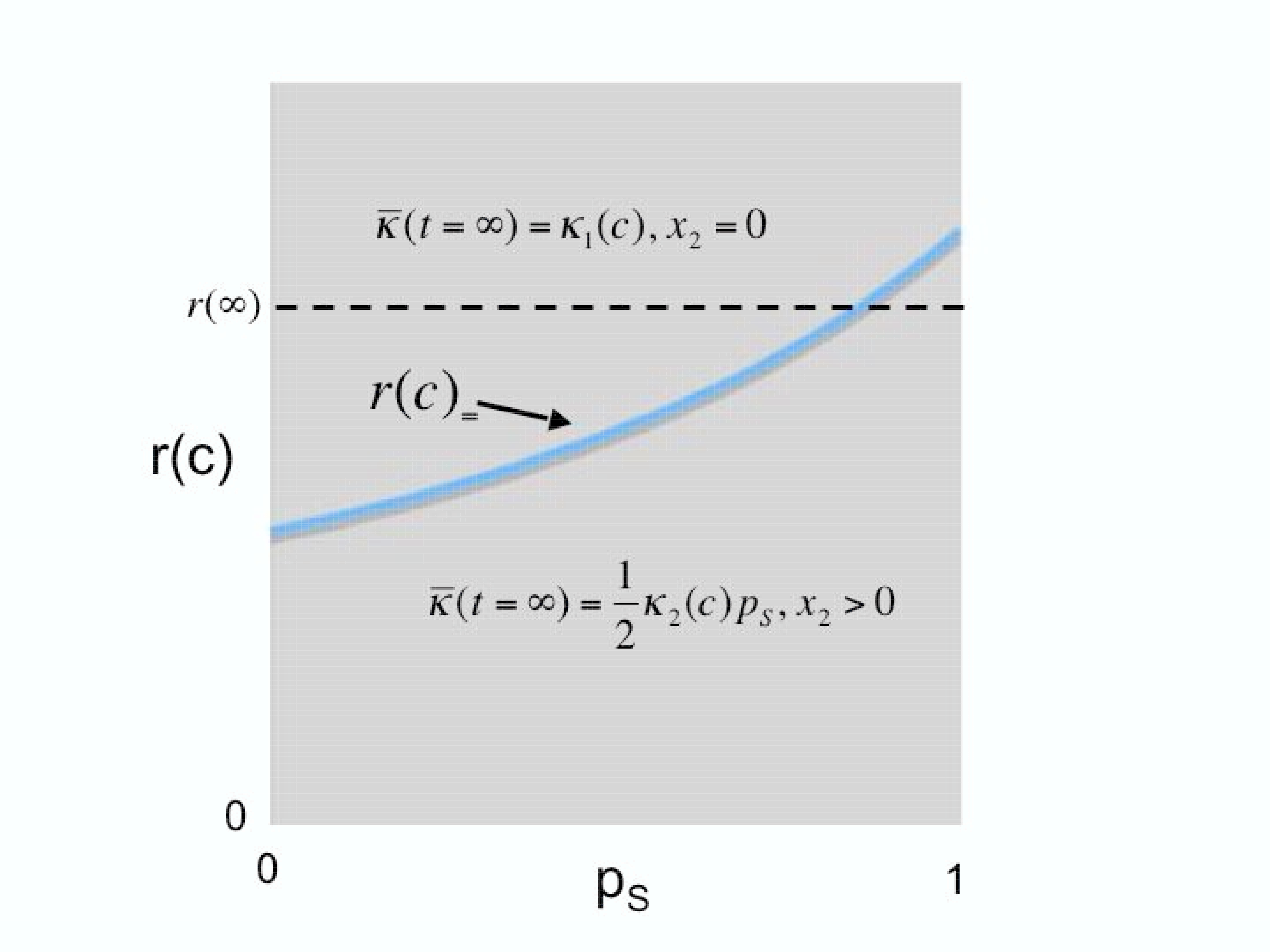}
\caption{Illustration of the two solution domains for our multi/single-cell
replication model.  Below $ r(c)_{=} $, the fraction of cells replicating via
the two-cell strategy is a positive fraction of the population.  Above $ r(c)_{=} $,
a localization to delocalization transition occurs over $ \sigma_S $, and the fraction
of cells replicating via the two-cell strategy drops to $ 0 $.}
\end{figure}

Figure 2 shows the various solution regimes as a function of $ r(c) $ and $ p_S $.  Note that,
for a given value of $ p_S $, there exists a low-concentration regime where the fraction of cells 
adopting the two-cell strategy is positive.  In this regime, there is a selective advantage for a 
genome to maintain a functional copy of the multicell switch $ \sigma_{S, 0} $.  At a critical 
concentration given by $ r(c) = r(c)_{=} $, resources are sufficiently plentiful that it becomes 
disadvantageous to instruct a cell to sacrifice its own reproductive ability for the sake of the other one.  
The reason for this is that, although the average fixed cost per cell is lower with the two-cell strategy, 
the cost of having to replicate the strategy outweighs the savings in fixed costs when resources are plentiful.  
Thus, once $ r(c) > r(c)_{=} $, the fraction of cells adopting the two-cell strategy disappears, and the population 
consists entirely of cells replicating via the single-cell strategy.

If $ r(c) > r(c)_{=} $ for $ p_S = 1 $, then varying $ p_S $ at this concentration
will never lead to a selective advantage for maintaining a two-cell survival strategy.  
If $ r(c) < r(c)_{=} $ for $ p_S = 1 $, but $ r(c) > r(c)_{=} $ for $ p_S = 0 $, then for 
sufficiently large $ p_S $ there will exist a finite fraction of the cells which replicate via the 
two-cell strategy.  As $ p_S $ drops below some critical value, denoted $ p_{S, crit} $,
the probability of incorrectly duplicating the strategy becomes sufficiently large that the fraction of 
cells replicating via the two-cell strategy disappears.  This concentration regime is interesting 
because it corresponds to a regime where replicating via the two-cell strategy is actually the 
advantageous one, but it might not be observed because of replication errors.  

Finally, once $ r(c) $ drops below $ r(c)_{=}|_{p_S = 0} $, then $ \kappa_1(c) < 0 $, so as
long as $ \frac{1}{2} \kappa_2(c) p_S > 0 $, there will exist a selective advantage
for maintaining the two-cell strategy in the population.  Due to mutation, this
will also lead to the maintenance of the single-cell strategy, although this
strategy is not self-sustaining in the population.

If, due to saturation, $ r(\infty) $ is finite, then one possibility is that
the parameters of our model are such that $ r(\infty) < r(c)_{=} $ at a given
$ p_S $.  Then for this value of $ p_S $, there will exist a selective advantage
for the two-cell strategy no matter what the external concentration of resource 
(the cells cannot metablize the resources sufficiently fast to eliminate the 
selective advantage for multicellularity).

The results of our model show that natural selection does not act on individual cells, but 
rather on the survival strategy as encoded for in $ \sigma_{S, 0} $.  Individual cells then are more 
properly viewed as vehicles by which the multicell strategy is passed on to the next generation.  When food 
resources become limited (or when the cells cannot rapidly metabolize the food resources present),
the effective growth rate of the multicell strategy is competitive with the total growth rate of
the single-cell strategies, resulting in its preservation in the population.  Essentially, it
becomes advantageous (from the point of view of the strategy) for several cells to pool their
resources together for the purposes of replicating a single cell.  When food becomes more plentiful, 
or when the rate of replication errors reaches a threshold value, the selective
advantage for retaining the strategy disappears, and delocalization occurs over the corresponding
region of the genome.

A potentially interesting avenue of future research is to determine whether there exist natural
bounds on the possible multicellular replicative strategies, and whether it is possible, using
thermodynamics and information theory, to connect these natural bounds to basic physicochemical
properties of the constituent reaction networks.  

\begin{acknowledgments}

This research was supported by the Israel Science Foundation.

\end{acknowledgments}


\begin{thebibliography}{10}
\expandafter\ifx\csname
natexlab\endcsname\relax\def\natexlab#1{#1}\fi
\expandafter\ifx\csname bibnamefont\endcsname\relax
  \def\bibnamefont#1{#1}\fi
\expandafter\ifx\csname bibfnamefont\endcsname\relax
  \def\bibfnamefont#1{#1}\fi
\expandafter\ifx\csname citenamefont\endcsname\relax
  \def\citenamefont#1{#1}\fi
\expandafter\ifx\csname url\endcsname\relax
  \def\url#1{\texttt{#1}}\fi
\expandafter\ifx\csname
urlprefix\endcsname\relax\def\urlprefix{URL }\fi
\providecommand{\bibinfo}[2]{#2}
\providecommand{\eprint}[2][]{\url{#2}}

\bibitem[{\citenamefont{Kreft and Bonhoeffer}(2005)}]{MULTICELL}
\bibinfo{author}{\bibfnamefont{J.U.}~\bibnamefont{Kreft}}
  \bibnamefont{and} \bibinfo{author}{\bibfnamefont{S.}~\bibnamefont{Bonhoeffer}},
  \bibinfo{journal}{Microbiology} \textbf{\bibinfo{volume}{151}},
  \bibinfo{pages}{637} (\bibinfo{year}{2005}).

\bibitem[{\citenamefont{Smith}(1998)}]{JMSMITH}
\bibinfo{author}{\bibfnamefont{J.M.}~\bibnamefont{Smith}},
  \emph{\bibinfo{title}{Evolutionary Genetics:  $ 2^{nd} $ edition}}
  (\bibinfo{publisher}{Oxford University Press},
  \bibinfo{address}{New York, NY}, \bibinfo{year}{1998}).

\bibitem[{\citenamefont{Britton}(2003)}]{BRITTON}
\bibinfo{author}{\bibfnamefont{N.F.}~\bibnamefont{Britton}},
  \emph{\bibinfo{title}{Essential Mathematical Biology}}
  (\bibinfo{publisher}{Springer-Verlag},
  \bibinfo{address}{London, UK}, \bibinfo{year}{2003}).

\bibitem[{\citenamefont{Tannenbaum and Shakhnovich}(2005)}]{QUASREV}
\bibinfo{author}{\bibfnamefont{E.}~\bibnamefont{Tannenbaum}}
  \bibnamefont{and} \bibinfo{author}{\bibfnamefont{E.I.}~\bibnamefont{Shakhnovich}},
  ``Semiconservative replication, genetic repair, and many-gened genomes:  Extending the
    quasispecies paradigm to living systems,''
\bibinfo{journal}{Physics of Life Reviews}, in press.

\bibitem[{\citenamefont{Wilke}(2001)}]{WILKE}
\bibinfo{author}{\bibfnamefont{C.O.}~\bibnamefont{Wilke}},
  \bibinfo{journal}{Nature} \textbf{\bibinfo{volume}{412}},
  \bibinfo{pages}{331} (\bibinfo{year}{2001}).

\end{thebibliography}
\end{document}